\documentclass{article}
\usepackage{graphicx}
\usepackage{amsmath}

\newtheorem{theorem}{Theorem}
\newtheorem{acknowledgement}[theorem]{Acknowledgement}

\input{tcilatex}

\begin{document}

\title{\textsc{Essay Review}\\
Classical Versus Quantum Ontology}
\author{P. Busch \\
Department of Mathematics, University of Hull\\
{\footnotesize Electronic address: P.Busch@maths.hull.ac.uk}}
\maketitle

\noindent D. Home, \emph{Conceptual Foundations of Quantum Physics: An
Overview from Modern Perspectives }(New York: Plenum Press, 1997), xvii+386
pp., ISBN 0-306-45660-5.

\noindent {\small ``\emph{The limits of my language} mean the limits of my
world.''}\newline
{\small (Wittgenstein, Tractatus Logico-Philosophicus, Proposition 5.6.)}

\bigskip

\section{Short review}

\noindent Quantum mechanics faces a strange dilemma. On the one hand it has
long been claimed to be an irreducibly \emph{statistical} theory, allowing
the calculation of measurement outcome statistics while being unable to
predict the behaviour of \emph{individual} microphysical processes. On the
other hand, quantum mechanics has been increasingly used, with stunning
success in the past few decades, to gain experimental control over
individual objects on an atomic scale. The old philosophical debates among
physicists over the interpretation of quantum mechanics have thus reached a
new stage where conceptual questions have obtained more precise formulations
and former \emph{Gedanken} experiments have been turned into actual
experiments. This situation has given an enormous boost to research into the
foundations of quantum mechanics, leading to a variety of promising
approaches towards a satisfactory theoretical account of individual
microphysical phenomena. It is not clear at present whether such an account
requires a modification of the standard quantum formalism or whether it can
be achieved within that formalism, on the basis of a consistent realist,
individual interpretation.

D. Home's book is devoted to contributing towards a clarification of this
question. It is evidently written by an inspired and established participant
in the ongoing quest to understand quantum mechanics and its description of
the physical world. The author openly admits to his own, Bohmian
ontological, preference; but in no place does the book become dogmatic about
this although that preference determines the line of.reasoning through most
chapters. It rather adheres to its motto, expressed beautifully in the
following quotations, chosen as the opening and closing words: \emph{`It is
the customary fate of new truths to begin as heresies and to end as
superstitions' }(p.ix), and: \emph{`The point is not to pocket the truth but
to chase it' }(p. 378)\emph{\ }The aim of the book is `to provide an
overview of the present status of the foundational issues of nonrelativistic
quantum mechanics. ... The \emph{need }to go beyond the standard
interpretation is a focal point of the book' (p. x). The book recommends
itself, appropriately, as just that: an \emph{overview} of conceptual issues
in quantum mechanics, from a \emph{physicist's }perspective. The physical
jargon and the style of presentation of mathematics are those of a
practising theoretical physicist; this will come easily to fellow physicists
but may in places provide difficulties for philosophers or mathematically
inclined readers.

The book focuses `firmly on the conceptual aspects, details of an
experimental or mathematical nature have been minimized wherever possible.'
With foundations of quantum physics being a substantial interdisciplinary
research field in its own right, it is impossible to give a comprehensive
account in one single book; hence any author is bound and justified to make
a choice of topics in accordance with his or her own expertise and
preference. Home's choice of perspective is that of a critical comparison
mainly of three modern approaches to solve central conceptual problems of
quantum mechanics -- the Bohmian causal approach, the decoherence models,
and the spontaneous wave function collapse models. A distinguishing feature
of the book is that this comparison is not so much concerned with the
relative theoretical or philosophical merits and difficulties faced by these
approaches, but that it focuses strongly on possible experimental
discriminations between them, taking into account modern technological
advances. Home's choice of topics (each one assigned a chapter) gives a fair
reflection of the core of current interest in foundational issues of quantum
mechanics: a review of the standard (Copenhagen) interpretation and the need
to go beyond it; the quantum measurement paradox; the classical limit of
quantum mechanics; quantum nonlocality; wave particle duality; quantum Zeno
effect; causality in quantum mechanics; and a reappraisal of Einstein's
critique of quantum mechanics. Each chapter starts with a superb
nontechnical, deeply reflected introduction of its topic, continues with a
lot of detailed, careful physical discussion, and concludes with an
extensive, valuable bibliography. Home has done a good job in trying to make
up for omissions by adding further relevant references. Overall, a list of
more than 700 works is provided, albeit quite strictly chosen within the
confines of the book's specific outlook. The index is arranged purely by
authors; a separate subject index would have been helpful in view of the
size of this volume.

The production quality of the book is acceptable, except for the display of
mathematical formulas; these look very much as if they were produced by
typewriters, and long formulas are especially hard to read. In addition,
careful proofreading could have helped to avoid a number of misprints in the
text and formulas. A curious example: on page 23, `...$g\left( t\right) $ is
switched on and off successfully...' rather than `successively'; and on page
247: `particle ontogonal approach' instead of \ `particle ontological
approach'. (Also, after several months of repeated study, the glue binding
gave in, leaving me with two loosely connected parts of the book in my
hands.)

Overall, I would recommend this book as a very valuable, up-to-date account
of quantum foundations from the perspective of a physicist interested in
possible experimental tests. It certainly provides a good graduate text for
students of physics seeking a deeper understanding of quantum mechanics, as
well as being a useful resource for researchers in the foundations and
philosophy of modern physics. The book complements in an original way
related recent publications on the foundations of quantum mechanics, such as
the texts on ``Bohmian'' quantum mechanics by D. Bohm and B. Hiley (1993)
and P. Holland (1993), or the ``strictly instrumentalist'' Quantum Theory
text by A. Peres (Peres, 1993).

Readers keen to get to know the book may stop reading here and return, if
they wish, to this review at a later stage. In the subsequent, more detailed
survey of the contents, I will point to some related lines of research
concerned with \emph{structural} aspects of quantum mechanics which are not
addressed in this book but awareness of which I believe is crucial for
obtaining a wider perspective on some problems. This attempt to place the
present book in a broader context is meant to illustrate the fact that in
the study of the foundations of quantum physics we seem to be facing a
diversity of research cultures which may benefit considerably from each
other if they could be brought into closer communication. It is only fair to
note that this point could be made with reference to any one of the books on
foundations of quantum mechanics, so that my remarks should not be
understood as criticisms of this particular book but rather as an
illustration of the general situation within this community.

It goes without saying that the choice of comments and issues raised is
again limited by a particular perspective, this time the reviewer's.\ If
Home's perspective is that of a theoretical physicist with a detailed
knowledge of the experimental side and a remarkable openness towards the
philosophical side, then I might characterise my own outlook as that of
someone trained as a theoretical physicist who ended up working `somewhere'
between theoretical physics and mathematical physics, while both of us seem
driven by an understanding of physics as \emph{natural philosophy, }or \emph{%
experimental metaphysics}. To obtain a balanced overview of the foundations
of quantum physics, it would\ indeed be desirable to complement and confront
the views presented in the book and in this review with those of a true
philosopher of physics, a professional mathematical physicist, and a real
experimental physicist. For the time being, with a symposium of that kind
outstanding for an indeterminate amount of time, I suggest that my
deliberations be accepted as one reader's dialogue with this book.

\section{Ontological position vs ontic indeterminacy}

Chapter 1 starts with a brief outline of the quantum formalism and its \emph{%
standard interpretation}. Home uses this term to refer to what he describes
as the common hypothesis of all versions of the so-called orthodox, or
Copenhagen interpretation: the hypothesis that a `wave function is
considered to be a complete description of the quantum mechanical state of
either an individual system or an ensemble of identically prepared systems'
(p. 16). The completeness claim, according to Home, `immediately implies
accepting an inherently statistical description in the microphysical
domain.' In the standard interpretation, the ``wave function''\footnote{%
I use `wave functions' in quotation marks in order to indicate that I regard
it as part of an outdated terminology. The term `state (vector)' would
suggest itself as a more neutral expression that does not carry the
connotation of a classical wave ontology.} is a representation of all
probabilistic knowledge about outcomes of possible measurements and as such
is devoid of any ontological content: `In other words in the standard
interpretation, the formalism of quantum mechanics or the quantum algorithm
does not reflect a well-defined underlying reality, but rather it
constitutes only knowledge about the statistics of observed results' (p.
17). Accordingly, Home rejects Bohr's and Heisenberg's interpretations of
the uncertainty relations to the extent that they go beyond the direct
experimental meaning in terms of spreads of measurement statistics: it
cannot be logically inferred from the uncertainty relation that individual
atomic particles could not have possessed simultaneous definite values for
noncommuting variables before any measurement.

These observations are made to indicate that there is scope for alternative
interpretations of quantum mechanics which provide more of a realist account
of phenomena in the domain of that theory. To substantiate this claim, Home
reviews Bell's critical analysis of von Neumann's influential
no-hidden-variables theorem as well as the Kochen-Specker theorem, to point
out the \emph{possibility} of contextual hidden variables. The \emph{need }%
to go beyond standard quantum mechanics derives from the infamous quantum
measurement problem, the classical limit puzzle, and the phenomenon of
nonlocality, issues to which the next three chapters are devoted.

Among the standard interpretations, Home distinguishes those which regard
the state vector either as a representation of an ensemble (of identically
prepared systems) or of an individual system. It is the latter which needs
to be contrasted with the realist interpretations. It is well known that an
unknown quantum state cannot be uniquely determined in a single run of a
measurement, and a simple argument is presented showing that any attempt to
nevertheless achieve this by means of state cloning must fail. However, Home
sketches an interesting proposal by Aharonov and Vaidman, known as \emph{%
protective measurement}, according to which the direct measurement of a
state on a \emph{single} system is possible provided enough is known about
that state so as to ensure that the state change due to the measurement is
negligible. This would indeed demonstrate that some objective reality can be
ascribed to the quantum state. In the simplest possible case, the proposal
would reduce to the situation where the state $\varphi $ is known and a von
Neumann-L\H{u}ders measurement of a simple observable, with outcomes
represented by the projections $P_{\varphi }=|\varphi \rangle \langle
\varphi |$, $I-P_{\varphi }$, would lead with certainty to the result
indicating $\varphi $. However, the protective measurement idea refers to
situations where the state need not be known in full, while at the same time
the known puzzles of individual state determinations are claimed to be
avoided. As Home makes repeated reference to protective measurements, it
seems worthwhile to point out a potential weakness of the existing proposals
that may restrict the validity of some of the implications suggested in the
literature.

The core feature of the various proposed schemes of protective measurements
is that for some quantum states, $\varphi $, (known to belong to a certain
class determined by the measurement interaction applied) it is possible to
obtain as a single reading the expectation value, $\langle \varphi |A\varphi
\rangle $, of some observable $A$, \emph{without }significantly changing the
state. If such a protective measurement is carried out for a sufficiently
large set of (noncommuting) observables $A,B,C,\dots $, then the values
obtained may suffice to infer what the state was -- and the system would
still be in that state. However, in these protective measurement schemes, no
analysis has been made of the magnitude of inference errors involved: it is
only shown that the \emph{expected }outcome is $\langle \varphi |A\varphi
\rangle $, but no estimate of a range of uncertainty has been given. In
fact, in a model of a joint measurement of position $Q$ and momentum $P$
very similar to one of the protective measurement schemes, it has been shown
that the realisation of the \emph{protective conditions }entails that the
likely error range for the joint values $\langle \varphi |P\varphi \rangle $%
, $\langle \varphi |Q\varphi \rangle $ is large compared to the
corresponding variances of momentum and position in the state $\varphi $,
respectively; hence a unique state inference is impossible (Busch, 1985;
Busch \emph{et al}, 1995, Sec. VI.3.2). It may be worth noting that the idea
of measuring the expectation value of an observable on a single system was
considered and critically examined in a rigorous spin chain measurement
model as early as 1978 (Zapp, 1978).

Whatever the value of protective measurements may ultimately turn out to be,
from a conceptual point of view the existence of von Neumann-L\H{u}ders
measurements of a discrete observable is sufficient to warrant the objective
reality of a pure quantum state as explained above. Apart from answering the
question of the epistemological or ontological status of the quantum \emph{%
state}, any interpretation of quantum mechanics will provide some rules
which determine the \emph{actual properties} of a system in a given state.
Home proceeds with an outline of the Bohmian ontological model, the best (if
not the only) elaborated nonstandard interpretation. In contrast to the
standard interpretations, the Bohmian model ascribes reality to a particular
physical quantity, the particle's position, in addition to the ``wave
function''. Home argues in several places throughout the book for the
necessity of distinguishing the kind of reality possessed by position from
the kind of reality possessed by the wave function: position is what we as
observers discern most directly, while the ``wave function'' makes itself
felt rather more indirectly, in its role as ``pilot wave'' guiding the
particle's motion. The ontological priority of position over the ``wave
function'' must be assumed, according to Home, if the Bohmian model is to
address adequately the measurement problem and the classical limit problem.
In line with this claim, Home maintains that since within the standard
approaches the ``wave function'' has a solely epistemic -- probabilistic --
function, it cannot provide an appropriate account of the emergence of a
definite, ontic (pointer) position of a macroscopic object as we know it
within the realm of classical physics.

However, the examples with which Home tries to illustrate the possible
superiority of the Bohmian over the standard interpretation are not entirely
conclusive (as he concedes). It is quite evident that the Bohm model allows
one easily to formalise barrier transmission and reflection times, or to
describe elementary particle trajectories as they are assumed in the
theoretical deduction of CP violation in kaon decay. Such concepts are
notoriously difficult to incorporate into quantum mechanics. Yet, a thorough
analysis of these experimental situations within standard quantum mechanics
is still lacking, and so it cannot be ruled out that a satisfactory and
rigorous account will finally be found. As recent studies seem to indicate,
the relevant tools are just about to be recognised: a formal representation
of time observables and particle trajectories within quantum mechanics can
indeed be given, namely in terms of positive operator valued measures
(a.k.a. POVMs) (see, e.g., Muga \emph{et al}, 2000). The uninitiated reader
may find an elementary introduction in the recent review of the monograph of
Busch \emph{et al} (1995) by Fleming (2000).

The description of quantum observables as POVMs can be seen as a completion
of the notion of observable within the Hilbert space framework of quantum
mechanics, just as the description of quantum states as density operators
constitutes a completion of the concept of states. In the latter case, the
state vectors represent the set of pure states, the extremal elements of the
full convex set of states, while in the former case, the traditional concept
of observables is included in the form of projection valued measures (PVMs).
It is quite conceivable that the extended (still standard) quantum language,
which comprises all \emph{effects}\footnote{%
I use this word in italic letters, to emphasise that it is a technical term.
In a measurement scheme, every outcome is represented by an \emph{effect},
which is determined as the unique operator whose expectation in each state
gives the probability of that outcome. Technically \emph{effects }are
positive operators bounded between the zero and identity operators $O$ and $%
I $. Operators $E,F$ are ordered as $E\leq F$ iff $\langle \varphi |E\varphi
\rangle \leq \langle \varphi |E\varphi \rangle $ for all vectors $\varphi
\in \mathcal{H}$. An operator $E$ is bounded between $O$ and $I$ iff $O\leq
E\leq I$.}, whether projections or not, can be given a consistent realist
interpretation of states and properties that incorporates ontic
indeterminacy. Rather than following Home's interpretation of Heisenberg's
potentiality concept in an epistemic sense, one may try to understand the
tendency to actualisation in an ontic sense. That is to say, it may be
possible to interpret the expression $\mathrm{tr}\left[ \rho \cdot E\right] $%
, which is usually taken to be the probability for the occurrence of the
outcome associated with the \emph{effect} $E$ if measured on a state $\rho $%
, as a measure of the degree of actualisation of the (sharp or unsharp)
property represented by $E$.

It is probably true that the feasibility of such an individual, \emph{%
unsharp (or fuzzy)} \emph{reality} interpretation of quantum indeterminacy
has not been sufficiently explored. But it seems to me that Home's argument
that the inference from the uncertainty relation to the ontological
indeterminateness of position and momentum, say, is not logically compelling
has a counterpart aimed at a Bohmian interpretation of position as a
definite property: Bohm's hidden variable theory is \emph{contextual }in the
sense that the measured values of (most) observables are bound to differ
from the possessed premeasurement values of the corresponding physical
quantities. Hence there is no compelling reason to \emph{assume }that
positions have definite values; and one may wonder what is gained by telling
stories about a physical system if what the stories tell is beyond
experimental control. One may add that the standard use of quantum mechanics
alone, without any recourse to hidden variables, has resulted in highly
sophisticated technologies enabling the control of and experimentation with
single micro-objects.

In discussing the interpretation of the uncertainty relations, Home points
out rightly that the variances of position and momentum in a quantum state
bear no logical connection with the measurement errors occurring in
simultaneous measurements of these quantities. Such a connection seems to be
suggested by Heisenberg in various semi-classical \emph{Gedanken}
experiments. A clarification of the question as to whether a measurement
uncertainty relation holds necessarily for position and momentum had long
been hampered due to the fact that no formal conception of joint
measurements for noncommuting quantities was available. However, there do
exist formal schemes for \emph{approximate} measurements of position and
momentum, and if the corresponding measurement couplings between the object
system and two probes are activated simultaneously, then the resulting
measurement scheme does constitute a joint approximate measurement of
position and momentum; and it turns out that however small the position and
momentum imprecisions are when the measurements are applied separately, the
joint coupling of both probes results in a readjustment of the individual
measurement imprecisions in such a way that they satisfy an uncertainty
relation. Furthermore, it follows within these models that upon obtaining a
phase space `point' reading, $(x,p)$, the quantum particle will be found
afterwards in a state in which position and momentum are \emph{unsharply }%
localised at that point, in the sense that the centers and variances of its
position and momentum wave packets are equal to the values $x$, $p$, and the
measurement imprecisions $\delta x$, $\delta p$, respectively. Details of
this rigorous `measurement imprecision' version of an uncertainty relation
can be found in Sec. VI. of (Busch et al, 1995), where it is also shown that
a convenient description of the measured joint position-momentum observable
can be given in terms of POVMs on phase space in such a way that the
measurement imprecision relation is automatically built in. The above
(tentative) indeterminacy interpretation of quantum uncertainties is thus
found to establish consistency between the possibilities of definition
(preparation of position and momentum values) and the possibilities of
determination (joint measurement of these quantities): what cannot be
prepared better than allowed by the preparation uncertainty relation cannot
be measured more precisely than allowed by the measurement imprecision
relation.

These considerations are intended to show that the development of a
mathematical theory of measurement and of observables represented as POVMs,
that took place alongside, and largely unnoticed by, mainstream theoretical
physicists, has opened up wider perspectives on some long-standing
conceptual issues and offered new possibilities of dealing with them. In
particular there is a well-developed theory of approximate joint
measurements of noncommuting quantities which permits the analysis and
interpretation of a variety of modern quantum optical and atomic
interferometric experiments and also guides the inception of new
experiments. Some important relevant contributions emphasising conceptual
issues are (de Muynck and Martens, 1990), (Appleby, 1998); for a more
extensive bibliography, cf. (Busch \emph{et al}, 1995).

While Home has demonstrated nicely in Chapter 1 that there \emph{is} scope
for going beyond the standard interpretation, the points raised above seem
to show that this may even be true in a sense not anticipated in the present
book. Besides the attempts to restore elements of a classical physical
ontology along the lines of the Bohmian model, there is, I believe, the
option of trying to develop a coherent `quantum ontology'; an emphatic
advocate of this route and of the ensuing need to develop and train
appropriate quantum intuitions and ways of thinking is J.-M.
L\'{e}vy-Leblond whose pleas for a progressive approach in incorporating
novel theoretical physical structures into our thinking about the world are
as mind-refreshing to read now as they were when they appeared (e.g.,
L\'{e}vy-Leblond, 1974, L\'{e}vy-Leblond, 1981; L\'{e}vy-Leblond and
Balibar, 1990). I agree with Home when he says that any approach will need
to be tested against its merits in dealing with the fundamental conceptual
problems of quantum mechanics. In Chapter 2 he turns to the `central riddle'
of quantum mechanics: the quantum measurement paradox.

\section{What is a measurement?}

In a quantum mechanical account of measurement processes, the object system
is brought into an interaction with an apparatus (or probe system) which
establishes an entangled state for the compound system. Thus, the notion of
a non-invasive measurement known from the realm of classical physics is no
longer an admissible idealisation in quantum physics. Starting with this
general observation, Home proceeds to sketch a variety of simple models in
which the entanglement is shown to arise as a necessary consequence of the
minimal requirement of a measurement to exhibit the eigenstates of the
measured observable. There is no general reflection on the meaning of
`measurement' in view of the fact that a quantum measuring process cannot be
said to reveal what is the case before the measurement, that is, to exhibit
the premeasurement value. (A discussion of this point can be found in (Busch 
\emph{et al}, 1991).) Hence we will adopt the minimal criterion suggested in
Home's models, namely, that a measurement should exhibit with certainty
which eigenstate the system was in, provided it was prepared in some
eigenstate. We may refer to this requirement as the \emph{calibration
condition}. The linearity of quantum dynamics then leads to the result that
for a superposition of eigenstates, the final state of system plus probe is
a superposition of product states corresponding to different pointer states.
Hence standard quantum mechanics does not seem to yield an account of the
occurrence of a definite outcome, represented by the apparatus being found
in one particular pointer state. This is the fundamental quantum measurement
paradox. Home immediately flags two important related problems, namely, the
question of the preferred pointer basis and the problem of explaining the
objective reality of outcomes in terms of changes of properties of the
macroscopic measuring device. The former is addressed by the decoherence
theory, while the latter leads up to the classical limit problem treated in
Chapter 3.

At this point it may be noted that the measurement paradox persists in the
case of measurements described by POVMs (Busch and Shimony, 1996). While it
can be expected that POVMs play a crucial r\'{o}le in the description of
macroscopic observables, there are indications that the conflict between the
unitary quantum dynamics and the occurrence of definite outcomes will not
disappear even if pointer observables are considered as unsharp quantities
(Busch, 1998, Del Seta, 1998).

Interestingly, Home's first example of a measurement coupling is one that is
purported to measure a continuous quantity, the position of a particle,
where the calibration condition can be satisfied only approximately. This
type of model is taken from von Neumann's book, and in the history of
quantum mechanics it has found applications in manifold variations. One may
therefore refer to it as \emph{the standard model }of quantum measurement
theory (Busch and Lahti, 1996). The essence of the model is that the initial
product of object state (position amplitude $\psi (x)$) and probe state ($%
\phi _{0}(y)$) is unitarily transformed into a correlated state, 
\begin{equation*}
\Psi _{0}(x,y)=\psi (x)\phi _{0}(y)\;\longrightarrow \;\Psi (x,y)=\psi
(x)\phi _{0}(y-x).
\end{equation*}
[Note: the summation sign in equation (2.1.12), p. 70 of the book under
review, should be removed.] If the probe amplitude is sharply peaked at $y=0$
initially, and if the object amplitude is sharply peaked at $x_{0}$, then
the final probe amplitude will be sharply peaked at $y=x_{0}$. This is the
approximate realisation of the calibration condition. It may be noted that
the probability of finding the probe in some interval can be expressed as
the expectation value of some positive operator in the initial object state.
This operator is a kind of smeared version of a position spectral
projection, and the family of all these operators associated with the
various subsets of position space constitutes a POVM, representing the
smeared position observable measured by this interaction. It is a bit ironic
that the first-ever and fairly realistic measurement-theoretic model of a
measurement is presented in the same book -- von Neumann's! -- that
introduces the highly influential, and highly idealised notion of a perfect,
repeatable measurement (of a discrete observable). The latter concept has in
fact had a damaging effect in that it was long considered almost as
synonymous with the term `measurement', while the former model could have
instantly led to the generalised notion of an observable represented as a
POVM and opened up a realistic approach to quantum measurement. For example,
as shown in (Busch and Lahti, 1996), this model lends itself most naturally
to the development of a theory of joint position-momentum measurements, with
the ensuing justification of the interpretation of the uncertainty relation
in terms of individual measurement imprecision sketched in the preceding
section.

The remaining part of Chapter 2 has three parts: a discussion of various
standard solutions and their inadequacies, a review of nonstandard
approaches, and a section offering some original experimental examples
devised to probe the relative merits of the nonstandard approaches. Among
the standard solutions, Home distinguishes the viewpoints of Bohr and
Heisenberg, the decoherence theory, and the Dirac-von Neumann projection
postulate. A common aspect of these approaches is the (implicit or explicit)
reference to the fact that the system (plus apparatus) is ultimately left in
a mixture of states corresponding to the relevant pointer states. Indeed,
Bohr's insistence on the need for a classical description of the measuring
devices may be read as an anticipation of the necessity to explain that
superpositions of macrosopically distinct (pointer) states are practically
never observed. Also, Heisenberg's notion of the (necessary) `cut' between
object and device (observer) seems to allow a formalisation in terms of the
partial trace operation over the device Hilbert space, which leaves the
system in a mixed state; the mobility of the location of the cut allows one
to perform the partial trace over the rest of the world beyond the device,
which leaves the system plus device in a mixed state. The quotations given
by Home make it apparent that Heisenberg considered these reduced density
operators to practically admit a subjective ignorance interpretation,
thereby anticipating the strategy of the decoherence approach. Decoherence
theories finally argue that the system (plus apparatus) is left in the
``appropriate'' mixture due to the ubiquitous interactions between the
device and its environment. Home stresses that it is our lacking an
explanation of the transition from the pure to the mixed state that
constitutes the measurement problem, and that reference to the practical
impossibility, due to decoherence, of distinguishing the two types of
description of the postmeasurement situation only constitutes a `FAPP'
solution.\footnote{%
`FAPP' is a famous acronym for `for all practical purposes', coined by John
Bell to characterise pragmatic attempts of dissolving fundamental quantum
problems.} Neither is a subjectivist, information theoretic account (Home
refers to Heisenberg and Zeilinger) satisfactory, according to which the
transition from the pure compound state to the mixture describes the change
of knowledge of the observers. As Home points out, this view does not do
justice to the fact that the change of knowledge is effected by a change in
the real state of affairs that this knowledge refers to.

Here it is interesting to note that the book contains no explicit discussion
of the two fundamentally different uses of mixed state (`density')
operators: Home gives a lucid account (already in Chapter 1) of the use of
density operators for the representation of statistical ensembles; however,
the fact that these operators arise as descriptions of subsystems of
compound systems emerges only rather implicitly at various points throughout
the book. It is important to note that a pure entangled state of a compound
system necessarily yields a mixed state description for each of its
subsystems, and that these density operators of the subsystems do not allow
an ignorance interpretation. A formal correlate to this is that every mixed
state has an infinity of possible convex decompositions into (generally
non-orthogonal) pure states, so that there is no \emph{a priori} preference
of a particular decomposition. These facts about state operators should be
an integral part in the teaching of the quantum formalism; but instead they
seem to be little known or noticed even in the expert literature. The
readers of this book should therefore be encouraged to follow up the
valuable references (10 and 11 of Chapter 1) given by Home.\footnote{%
The mathematical feature of the non-unique convex decomposability of any
mixed state was first observed as a general phenomenon by Schr\"{o}dinger
who was also fully aware of the interpretational implications (Schr\"{o}%
dinger, 1936). A complete characterisation of the possible decompositions of
a non-pure density operator has been given in (Hughston \emph{et al}, 1993)
for the finite-rank case, and in full generality in (Cassinelli \emph{et al},
1997) , based on a fundamental partial results of (Hadjisavvas, 1981).} A
weakness of the decoherence account is that it does not recognise the need
to address the fundamental nature of this distinction.

If the standard framework of quantum mechanics is unable to resolve the
measurement problem, it follows that alterations, either of the formalism or
of the interpretation, have to be taken into consideration. The nonstandard
approaches considered at some length by Home are: the many-worlds
interpretation, the Bohmian model, and the dynamical models of spontaneous
`wave function' (state vector) collapse. Very clear and valid critical
assessments are given of the merits and difficulties of each of these
approaches; in particular, some of the dynamical collapse models are
presented in considerable technical and quantitative detail, in preparation
for the subsequent exploration of possible experimental discriminations
between these models, and the Bohmian model and the decoherence theoretical
accounts.

At this point the reader may wonder whether it would be possible to give a
systematic overview of the different interpretational options. I believe
this question can be answered in the positive. If the standard rule of
associating definite values with eigenstates is regarded as the root of the
measurement problem, then this would suggest a systematic investigation of
possible alternative interpretational rules specifying which quantities can
be regarded as having definite values in relation to a given state. A
comprehensive classification of such \emph{no-collapse }interpretations has
indeed been achieved in the recent book by J. Bub (1997), on the basis of a
seminal paper by Bub and Clifton (1996). This work allows one to consider
early and very different interpretations such as Bohr's and von Neumann's
under a common perspective with the Bohmian model, the many-worlds and other
more recent variants such as modal interpretations and consistent histories
accounts. From the point of view of such a general analysis it becomes clear
that the decoherence theory does not constitute a stand-alone approach to
the measurement problem, unless it is supplemented with further
interpretational commitments. In other words, decoherence is a physical
phenomenon that plays a crucial part in the physics of macroscopic systems,
and as such it figures in probably all no-collapse interpretations of
quantum mechanics.

On the other hand, if one wishes to hold on to the eigenvalue-eigenstate
rule, the only option left seems to be that of modifying the standard
quantum formalism. There are basically two ways of removing the undesired
superpositions of pointer states (or of more general macroscopically
distinct states). The projection postulate, either in Dirac's stochastic
quantum jump version or in von Neumann's collapse form can apparently be
realised in two ways only: either one modifies the dynamical law of quantum
mechanics so that for isolated microsystems the Schr\H{o}dinger equation is
valid as a good approximation while for large systems a stochastic
contribution in the generalised dynamical equations becomes predominant,
leading to spontaneous destructions of coherent superpositions; or one makes
room for the existence of superselection rules and the associated classical
observables. But since these are known to emerge naturally only in the
context of quantum systems with infinitely many degrees of freedom, the
standard formalism of the Hilbert space quantum mechanics of finitely many
particles would have to be extended to allow a modelling of macroscopic
objects as infinite systems in an ontic sense. The extended framework is
that of quantum field theory or its abstract version, $C^{\ast }$-algebraic
quantum theory.

Home addresses only very briefly one version of the superselection rules
approach, the many-Hilbert spaces theory. Therefore two supplementary
remarks may be in place. A substantial philosophical assessment of the
algebraic theory of superselection in relation to the quantum measurement
problem can be found in (Landsman, 1995). The $C^{\ast }$-algebraic quantum
formalism provides a natural framework for rigorous studies of decoherence
as well as spontaneous dynamical collapse mechanisms, as indicated in recent
work in the context of quantum filtering theory (Belavkin, 1994; Belavkin
and Melsheimer, 1995).

The final topic of Home's chapter on measurement is a very interesting
discussion of two experimental proposals. The first is a neutron
interferometry experiment potentially capable of discriminating between
decoherence effects and spontaneous wave function collapse models. Suitable
variations of parameters in the various models show that environment-induced
decoherence and spontaneous collapse lead to quantitatively different
predictions regarding the occurrence or non-occurrence of observable
interference effects. In the second, more tentative and less detailed,
proposal, a DNA molecule is considered as a detection device to probe
ultraviolet photon emission from a source. The mesoscopic nature of such
macromolecules suggests that their functioning as quantum probes does not
prevent them from assuming either one of their macroscopically distinct
states rather than persisting in an entangled superposition state with the
photon.

Home points out some quantitative difficulties met by the present
decoherence and collapse models in an attempted explanation of such an
experiment. In the case of the Bohm model, Home speculates that the
ontological definiteness of position may allow the conclusion that
interference effects of the -- possibly overlapping -- ``wave functions'' of
the uv-damaged and undamaged DNA configurations are suppressed
(unobservable). This would again demonstrate that `within\ the Bohmian
scheme, ontological position must be ascribed a more fundamental reality
than the wave function' (p. 132). Home's main conclusion drawn from these
examples is that experiments of the kind described here may provide useful
hints for further refinements of the nonorthodox approaches.

Yet, while it is true that specific models of decoherence and spontaneous
localisation, say, may lead to distinguishable experimental predictions, it
is unclear whether an interpretation based on dynamical collapse can be
experimentally discriminated from the \emph{no-collapse }interpretations.
For in the latter it is just the \emph{meaning }of the term `objective
reality' that has been changed; in fact, any no-collapse interpretation can
adopt some form of stochastic dynamic model as an \emph{effective}
description of the state of affairs. It is thus an open question whether a
change of interpretation (concept of reality) will suffice for a resolution
of the measurement problem, or whether a change of theory (dynamics;
superselection rules) is ultimately needed. At present, every researcher
into this problem must be prepared to make a choice and a long-term
commitment to one of these options and pursue its elaboration, without being
able to know whether this will ultimately lead to its confirmation or
refutation. The question to be answered by any no-collapse interpretation is
whether it affords a coherent reformulation of the bulk of theoretical
results of quantum physics in its own terms; in particular, it needs to
provide a satisfactory account of the selection of its specific `preferred'
observable (Bub, 1997). The central task for the spontaneous collapse
theorist is to go beyond the exploratory stage of \emph{ad hoc} models and
establish a general axiom of modified quantum dynamics that can be accepted
as compelling and aesthetic as the unitary Schr\H{o}dinger evolution has
always been regarded to be.

The involvement of large objects in quantum measurements connects the
measurement problem with the broader question of the classical limit of
quantum mechanics. The nonstandard approaches discussed so far take for
granted the universal validity of quantum mechanics or one of its
modifications. Home refers to Leggett and (`even') Feynman as two
distinguished authorities who pointed to the possibility that quantum
mechanics could fail for large objects. The expectation is that before long
experimental technology may have advanced far enough so as to allow us to
test this possibility. Hence a good understanding of the relationships
between the quantum and classical descriptions of macroscopic systems will
be required. This is the subject of Chapter 3.

\section{Is there a classical limit of quantum mechanics?}

To begin with, I would like to cite another distinguished authority whose
life work constitutes an essentially negative answer to this question. For
the briefest summary of G. Ludwig's monumental studies of the quantum theory
of macrosystems seems to be that quantum mechanics is not easily capable of
providing the objective description appropriate to the behaviour of
macrosystems. Ludwig concludes that a hierarchy of theories is needed to
account for the whole range of phenomena from the microscopic to the
macroscopic realms. In fact, in his approach the Hilbert space quantum
mechanics of microsystems is first deduced from \ the objective description
of macroscopic devices (Ludwig, 1985). The view that the extrapolation of
quantum mechanics into a many-particle theory of macrosystems is a more
comprehensive theory than the objective description of macrosystems has,
according to Ludwig, `generated unsurmountable difficulties for explaining
the measuring process'. Accordingly, his solution of the measurement problem
is based on the construction of an objective theory of macrosystems that is
more comprehensive than the extrapolated quantum mechanics (Ludwig, 1987).
The compatibility between these two theories is expressed as the fact that
approximate embedding maps between them can be formulated. Research along
the lines of this programme is being pursued by the Marburg and Milan
groups, e.g\emph{.},. (Lanz, 1994), (Lanz and Melsheimer, 1993)).

It is reassuring to observe that the general conclusions obtained from such
a comprehensive \emph{structural} \emph{study} of the quantum-classical
relationship are in agreement with the implications drawn from the very
concrete \emph{case studies} of classical limit procedures presented by
Home. On reading his Chapter 3 it becomes evident that there are many
strands to the classical limit problem, including the traditional
semiclassical methods ($h\rightarrow 0$ limit, large quantum number limits,
Ehrenfest's theorem) as well as the decoherence models and the Bohmian
approach. In each case it is shown in detailed, explicit examples that
certain classical features of large systems can approximately be described
in terms of quantum mechanics; but it is also made clear that there remains
room for genuine quantum effects in large systems. Thus there is scope for
future experimental tests of this extrapolation of quantum mechanical
predictions into the macroscopic realm. But Home points out that before any
definite conclusions can be drawn, many more, and more realistic, case
studies will have to be delivered within each of the approaches discussed.

Despite the unquestionable merit of this chapter as an introductory survey
of the classical limit problem, it is crucial to point out that a whole
range of important contributions (in addition to Ludwig's work) is left
unnoticed. For example, the demand for more realistic modelling, including
the development of a range of theoretical tools, had already been met to a
significant degree in the case of the decoherence approach, by a book that
appeared almost simultaneously with the present one: namely, (Giulini \emph{%
et al}, 1996), reviewed in (Donald, 1999). Other contributions are concerned
with the \emph{structural} similarities and differences between quantum and
classical mechanics, as opposed to the quantitative aspects considered in
the present book. An excellent up-to-date exposition of mathematical aspects
of quantisation theory can be found in (Landsman, 1998) (cf. the forthcoming
review by G. Emch in this journal), while (Schroeck, 1996) (reviewed in
(Landsman, 1999)) approaches the quantum-classical relationship from the
point of view of quantum mechanics on phase space.

Home's analysis of the classical limit is centered on the following three 
\emph{classicality }criteria: (1) the time evolution of a macroscopic system
should (approximately) be describable in terms of a classical dynamical law
(for its relevant state variables); (2) a macrosystem should be described as
an object that is well localised at all times; (3) a macro-object can be
measured non-invasively, that is, without affecting the outcomes of
subsequent measurements. As regards, these criteria, one may wonder why only
position should be required to have a definite value. The dynamics of a
classical `particle' is deterministic with respect to position and momentum
taken together as the state variables, not with respect to position alone.
Similarly, non-invasive measurability should be stipulated to hold for \emph{%
all} macroscopic quantities, not position alone. Hence one is forced to
confront the fundamental structural difference between the quantum and
classical description of a particle; the question to be asked is: how can
the familiar deterministic phase space description of macroscopic classical
particles be extracted from the quantum mechanical Hilbert space
description? Attempts to resolve this problem have led to a variety of \emph{%
phase space}\ formulations of quantum mechanics, ranging from the
Wigner-Weyl formalism and Husimi distribution to geometric and other phase
space quantisation schemes. Powerful mathematical and theoretical tools for
the treatment of foundational as well as concrete quantum mechanical
problems have been developed in these approaches, leading to valuable
structural insights into the problem of quantum-classical compatibility.
Yet, a coherent, generally accepted account of what exactly constitutes the
classical limit of quantum mechanics is still lacking; and it may not be
achieved without realistic case studies of macroscopic systems which make
full use of the existing conceptual tools. In particular, such studies
should make explicit the macroscopic nature of the systems, that is they
should manifestly take into account the large number of degrees of freedom
of these systems. Interesting approaches where `macroscopic' is explicated
using the tools of nonstandard analysis (i.e., considering Planck's constant 
$h$ as infinitesimal) have led to a structural transition from quantum to
classical descriptions; (Werner and Wolff, 1995), (Ozawa, 1997).

In summary, to date the question whether Giulini et al or Ludwig is right --
that is the question whether quantum mechanics does or does not suffice to
explain the emergence of a classical world in the macrodomain -- must be
regarded as largely open: the starting points of \ these approaches are so
different in their philosophical outlook and ensuing conceptual elaborations
that a confrontation of their contrasting conclusions will require extensive
further investigation. Examples of extensive recent studies complementing
the material of Home's Chapter 3 are the books of Landsman (1998) and
Schroeck (1996) mentioned above, as well as Stulpe (1997), which provide
good starting points for a systematic treatment of these questions. A very
surprising perspective on quantum mechanics, displaying its striking
contrasts with classical mechanics in a novel way, has been discovered and
developed during the 1990s by Beltrametti and Bugajski (e.g., Beltrametti
and Bugajski, 1995), who introduced a classical extension of quantum
mechanics in which quantum states are represented as mixed classical states
and quantum \emph{effects} are represented as fuzzy classical \emph{effects}%
. This example shows that we should not expect that every possible way of
confronting quantum mechanics with classical ontological ideas such as
hidden variables has already been explored or even envisaged. The vast
``distance'' to be passed in the transition from quantum to the classical
(or conversely) is strikingly illuminated in a philosophical case study of
elementary particle tracks, which coincidentally enhances Ludwig's position
regarding the necessity of a chain of theories linking the accounts of the
microscopic with those of the macroscopic (Falkenburg, 1996).

\section{Quantum nonlocality, superluminal signals,\protect\linebreak\ and
all that}

Chapter 4 on `Quantum Nonlocality' provides a careful explanation of what
constitutes a nonlocal effect. Home distinguishes between two types of
nonlocality -- kinematic and measurement-induced. The former kind is
exemplified by the famous Einstein-Podolsky-Rosen experiment and is
generically represented by pairs of spatially separated systems in entangled
states. Home reviews the attempts to provide local realistic accounts of a
variety of situations and the ensuing Bell-type inequalities or relations
without inequalities, which are in conflict with both quantum mechanics as
well as in some cases with actual experiments.

Interesting novel points discussed in great detail are the quantum
mechanical predictions of violations of local realism even in the
macroscopic limit and an experiment exhibiting nonlocality in single-photon
states. Instances of measurement induced nonlocality arise in correlated,
spatially separated systems if the collapse of the state vector is taken as
an objective real process occurring in the individual case: this is nicely
demonstrated in a model-independent example, thus reinforcing the notion of
`objectification at a distance' or `passion at a distance' (Shimony, 1984).
Other intriguing instances of this type of nonlocality involve
negative-result measurements and the novel process of quantum teleportation
(which since the publication of the book has been experimentally realised).
Home emphasises that quantum nonlocality does not necessarily involve
spacelike separations and thus relativistic considerations. Going somewhat
beyond the scope of the book (nonrelativistic quantum mechanics), a brief
discussion of the problem of spacelike nonlocalities and the ensuing
`danger' of superluminal signalling, and hence violations of relativistic
causality, is given.

In this context, the issue of individual state determinations becomes
crucial, which has been raised in various places in the book: the
possibility of protective measurements, and the impossibility of state
cloning (pp. 20-23). Home reviews the proposal that individual state
determination would be feasible \emph{if} it were possible \ to measure
non-Hermitian\ operators with their non-orthogonal systems of eigenstates,
and emphasises that this would enable superluminal signalling using EPR
entangled systems. What seems to be lacking in the relevant literature is
any attempt to develop a theory of measurements of such operators. It seems
to me that the only conceivable route to making operational sense of such
proposals is by way of the standard measurement formalism and the ensuing
POVM approach. After all, the non-Hermitian operators in question are
associated with a POVM in the same sense as a standard self-adjoint operator
is associated with its spectral measure. There do exist general results to
the extent that state cloning or other ways of discriminating non-orthogonal
states using measurements involving POVMs are equally doomed to fail as was
the case with standard observables (see, e.g., (Busch, 1997)).

If one enters the domain of relativistic quantum mechanics, the issue of
nonlocality assumes an entirely new level of complexity: the definition of a
local or nonlocal phenomenon must be based on a precise concept of \emph{%
localised }processes or operations; and the known ways of formalising
localised states or localisation observables lead to implications that seem
to be in conflict with relativistic causality. For recent reviews of the
conceptual aspects involved, cf. (Butterfield and Fleming, 1999), and
(Busch, 1999). Even the \emph{definition} of (sharp, i.e. PVM) position
observables for elementary systems is limited to the case of massive
particles or massless particles of spin less than 1. It is only within the
extended set of POVMs that a unified account of relativistic particle
localisability can be achieved, namely, in terms of covariant phase space
observables (e.g., (Schroeck, 1996), (Brooke and Schroeck, 1996)). In the
current discussions of nonlocality, the localisation of the measurement
operations involved is always tacitly assumed but apparently there is no
attempt to make this assumption formally explicit. Hence a coherent account
of these phenomena in terms of relativistic quantum theory is still waiting
to be carried out. This becomes even more urgent in view of recent
experimental demonstrations of (i) EPR-type nonlocality with entangled
photons at distances of more than 10km (Zbinden \emph{et al}, 2000) and (ii)
photons tunnelling through opaque media with ``superluminal'' speeds
(Cologne, 1998).

\section{Complementarity \emph{versus }Uncertainty?}

In Chapter 5, Home discusses `Wave particle duality of light and
complementarity'. After a critical review of Bohr's views and some
traditional early formalisations of the idea of complementarity, a variety
of modern quantum optical experiments are described, concluding with the
provocative suggestion that complementary wave and particle aspects can
coexist, after all, possibly in some contrast to Bohr's intuitions. The
chapter concludes with a careful examination of the empty-wave paradox as a
difficulty of the Bohm and de Broglie causal theories. One may wonder
whether a radical alternative approach to the whole issue would be to
abandon the ``wave'' and ``particle'' terminology, along with undertaking a
fundamental revision of the underlying ontology. It may be noted that
conclusions similar in some sense to Home's, regarding the coexistence of
information about path observables (``particle'' properties?) and
interference observables (``wave'' properties?) have been obtained in
measurement theoretic analyses of similar experiments. It is in fact
possible to formalise the notion of a joint approximate measurement of such
pairs of `complementary' observables. The complementarity is then expressed
in the reciprocal behaviour of the degrees of precision available at the
same time (see, e.g., (Martens and de Muynck, 1990a,b), (de Muynck \emph{et
al}, 1991), (Busch \emph{et al}, 1995)).

The idea of complementarity in quantum mechanics, however vague its
descriptions by Bohr may have been regarded, has been a source of
inspiration in the search for appropriate interpretations of quantum
mechanics. This is well exemplified by Home's Chapter 5. Yet it is rather
disturbing to see that various strands of important investigations have
remained largely unnoticed. There have been confused debates about the
logical relations between complementarity and uncertainty `principles' ever
since the quantum pioneers introduced these notions. The style, and the
conceptual and formal level of these discussions have advanced surprisingly
little beyond the original works from the 1920s and 1930s. It is amazing to
see that despite this conceptual obscurity, some very fascinating novel
experimental realisations of former \emph{Gedanken} experiments illustrating
complementarity have been conceived and carried out. One recent example is
an atomic interferometric demonstration of a link between complementarity
and entanglement and of the fact that dynamic disturbances cannot (always)
be made responsible for the destruction of interferences in `which path'
experiments (D\H{u}rr \emph{et al}, 1998). The controversy in the journal
`Nature' on `complementarity \emph{versus} uncertainty' leading up to this
experiment (see the references in (D\H{u}rr \emph{et al}, 1998)) could
probably have been cut short by taking into account existing relevant
studies on the subject.

To begin with, `complementarity' and `uncertainty' are not (any more)
`principles' on which the presentation and teaching of quantum mechanics are
(to be) based. They are more appropriately regarded as logical consequences
of the formalism. As such, their logical relation cannot strictly speaking
be investigated \emph{within} the Hilbert space framework. Such an analysis
requires a more general theoretical framework in which both ideas can be
formulated as contingent postulates. Only then can the question be asked
whether or not one implies the other, or whether or not they both have some
common implications. Answers can be found in (Lahti, 1980; 1983). Next,
uncertainty and complementarity can be understood as relations (between
observables) within standard quantum mechanics; even then there are
different possible formalisations. We have discussed the case of the
uncertainty relations in an earlier section. Valuable studies of
quantitative aspects of complementarity and uncertainty relations can be
found in (Lahti, 1987), (Martens and de Muynck, 1990a,b), or (Uffink and
Hilgevoord, 1985).

\section{The Quantum Zeno effect and time as an observable}

The last strictly physical chapter of the book deals with the quantum Zeno
effect -- the fact that under certain conditions the dynamical evolution of
a quantum system can be inhibited by continuously observing it. For example,
continuous monitoring of an unstable state may have the effect of `freezing'
the evolution altogether. This phenomenon is paradoxical if one ignores the
fact that in quantum mechanics, and contrary to classical physics,
measurements can \emph{not} be regarded as non-invasive. Home reviews simple
models of continuous observations of decaying systems where the effect
depends on deviations from the exponential decay law in the short time range
which are theoretically required but not yet experimentally exhibited. This
is followed by a careful analysis of the famous `quantum telegraph'
experiment of Itano \emph{et al}, showing that it demonstrates an
interaction-induced -- as opposed to measurement-induced -- inhibition of
transitions. Home then describes the status of a variety of ingenious
experimental proposals which can be expected to eventually lead to
conclusive tests of the quantum Zeno effect.

The central practical difficulty is one of making rapid sequences of
measurement, with an enormously high degree of temporal resolution, within
the order of the lifetimes of the observed systems. Another, conceptual,
issue not raised is the question as to whether a `continuous observation' is
adequately modelled as a rapid sequence of ordinary (von Neumann-L\H{u}ders)
measurements. There may be a subtle but fundamental difference between the
experimental approaches toward answering the two questions: `Has the system
decayed yet at time $t_{n}\in \left\{ 0,t_{1},\dots ,t_{N}=T\right\} $?',
and `When did the system decay during the period $\left[ 0,T\right] $?' In
the first case the answer will be sought by making a yes-no measurement of
the simple observable $\left\{ P_{\psi },I-P_{\psi }\right\} $, where $%
P_{\psi }$ is the projection onto the unstable initial state of the system.
In the latter case the experiment consists of placing detectors around the
system and \emph{waiting} for its decay products to show up. This addresses
the question about the time of the occurrence of an event, considered as an 
\emph{event time observable}. A quantum theory of time measurements is
largely still waiting to be developed, although the POVM approach has
provided some promising modelling, primarily of photon counting processes in
quantum optics. For a review, cf. (Srinivas and Vijayalakshmi, 1981),
(Srinivas, 1996). Since the appearance of the present book, questions such
as `When does a measurement occur?', and `When does a particle (decay
product) pass a certain space region (detector)?' have become the subject of
renewed intense interest in experimental and theoretical physics, leading to
an increased awareness and appreciation of the deep open conceptual problems
involved (Muga and Leavens, 2000).

\section{Causality, reality, objectivity}

The last two chapters are devoted to philosophical issues. Chapter 7
presents a discussion of the possible meanings of causality and its status
in classical and quantum mechanics. An assessment is given of the various
different ways in which the standard interpretation, the Bohmian model and
the dynamical collapse theory attempt to cope with the apparent
indeterminism of individual measurement outcomes. The attitude of the
standard approach is characterised as a resignation to accept acausality at
the individual level and to be content with the validity of causality at the
statistical level (of the evolution of probabilities). By contrast, the
Bohmian causal theory offers a way of restoring a manifest causal link
between pre- and post-measurement situations -- even if at the expense of
having to acknowledge that this causal account is \emph{inaccessible }to
observation as a matter of principle. The dynamical collapse theories
finally cast the indeterminism of stochastic jumps occurring in any
measurement into the form of a law, thus providing a logical basis for the
concept of statistical causality. The tensions with relativistic causality
faced by the latter two approaches are briefly explained, noting that both
of them have their ways of evading causal paradoxes, either by accepting a
preferred frame of reference or by careful selection of the collapse
dynamics, respectively.

The \emph{r\'{o}le }of the concept of causality in the complex process of 
\emph{constituting objective experience} is not addressed. Home comes close
to such a consideration in the final chapter where he offers a reappraisal
of Einstein's case for realism. On the basis of a number of carefully
selected and well placed quotations from Einstein and his contemporaries,
Home describes Einstein's turn from his early positivistic preferences to
his ultimate advocacy of \emph{local, causal realism}. In trying to exhibit
Einstein's motivation for his strict adherence to locality, Home makes the
following key observation (pp. 367-368): `Einstein's point was not that
nonlocal actions are inconceivable but that their existence undermines
physical science. If distant nonlocal influences are permitted, then unless
these are eliminated ..., we cannot trust measurement results to indicate
that a system is in a specific state, possesses specific properties, and so
on... Thus Einstein believed that the locality condition was necessary to
ensure the existence of closed systems and therefore the possibility of
testing theories:

\begin{quote}
{\small if this axiom were to be completely abolished, the idea of the
existence of (quasi) enclosed systems, and thereby the postulation of laws,
which can be checked empirically in the accepted sense would become
impossible.}
\end{quote}

Note that the primary motivation behind Einstein locality is not the
relativistic requirement that no signal may propagate faster than light but
rather a more general consideration related to a fundamental methodological
principle of physical science.'

Home here touches upon an issue that is central to one particular approach
towards reconstructing quantum theory: the Cologne version of `quantum
logic'. Quantum logic was initiated by von Neumann and Birkhoff who analysed
the proposition structure entailed by the lattice of subspaces of Hilbert
space. Various researchers considered quantum logic as a revision of
classical logic which allowed one to maintain the value definiteness of
propositions without running into the contradictions that classical logical
rules would otherwise lead to. Later it was realised that the \emph{%
aprioristic }structures of logic could be recovered by a transcendental
philosophical argumentation much in the same way as was carried out by Kant
for epistemological categories such as substance and causality. This led to
a reconstruction of the quantum language via a reflection on necessary
conditions of the accepted form of a scientific language about \emph{object}%
-ive scientific experience. A convenient form for this programme was
provided by the theory of dialogue games (Mittelstaedt, 1978), (Stachow,
1980). Moreover, it has been possible to exhibit specific features of the
quantum mechanical proposition lattice -- atomicity and covering law -- as
consequences of the condition that this language refers to \emph{individual }%
objects (Stachow, 1985).

On the formal side, a central aim of the quantum logics approach was the
derivation of the Hilbert space realisation of the proposition lattice from
physically motivated assumptions. In fact, as is well known, any irreducible
orthocomplemented, orthomodular lattice of chain length greater than 3 can
be identified with a (sublattice of a) lattice of subspaces of some
orthomodular vector space, in such a way that the associated bilinear form
of that space determines the orthogonality relation. The construction of the
isomorphic embedding in question fixes uniquely the skew field over which
the vector space is defined. It was long believed that the only candidate
fields for which this construction worked are the `classical' fields of the
real or complex numbers, or the quaternions. However, in 1980, examples of
`non-classical' orthomodular vector spaces were discovered, and the whole
quantum logic programme stalled for about 15 years as it was not clear
whether a lattice theoretic property could be formulated that would select
the `classical' fields. Such a condition was indeed found in the mid-1990s,
leaving still open the problem of a physical motivation. For a survey of
this development, cf. (Holland, 1995).

The idea of deducing quantum structures from conditions of objective
experience suggests a thorough revision of the Kantian programme in order to
examine whether this approach can be appropriately adapted so as to
encompass modern physical theories. Such a project would not only examine
the role of properties such as locality in the constitution of objects but
would include all other categories originally proposed by Kant, including
causality. The first steps into this major philosophical enterprise have
been taken within the Cologne group (Mittelstaedt, 1986), (Strohmeyer,1987,
1995), but much work remains to be done, particularly in exhibiting the
implications of the philosophical findings for the interpretation of quantum
mechanics. At the present stage, I feel, there is room to hope that a
quantum ontology can be formulated which gives a framework for a viable
alternative to the Bohmian model, namely, in the form of a coherent
`indeterminacy', or `unsharp reality', interpretation of quantum mechanics
as a theory of individual objects (Busch et al, 1995), (Mittelstaedt, 1995).

\begin{acknowledgement}
My sincere thanks go to Marcus Appleby, James Brooke, Jeremy Butterfield,
and Chris Shilladay for their critical (proof)reading and/or numerous
valuable comments and suggestions on a draft version of this work.
\end{acknowledgement}

\bigskip 

\noindent {\Large References}

\bigskip

\noindent Appleby, D.M. (1998) `Concept of experimental accuracy and
simultaneous measurements of position and momentum' \emph{International
Journal of Theoretical Physics }\textbf{37}, 1491-1509.

\noindent Belavkin, V.P. (1994) `Nondemolition Principle of Quantum
Measurement Theory', \emph{Foundations of Physics }\textbf{24}, 685-714.

\noindent Belavkin, V.P., and Melsheimer, O. (1995) `A Hamiltonian Solution
to Quantum Collapse, State Diffusion and Spontaneous Localisation', in \emph{%
Quantum Communications and Measurement}, eds. V.P. Belavkin, O. Hirota, and
R.L. Hudson (New York: Plenum), pp.201-222.

\noindent Beltrametti, E., and Bugajski, S. (1995) `A classical extension of
quantum mechanics', \emph{Journal of Physics }\textbf{A28}, 3329-3343.

\noindent Bohm, D., and Hiley, B.J. \emph{The Undivided Universe: An
Ontological Interpretation of Quantum Theory }(London: Routledge).

\noindent Brooke, J.A., and Schroeck, F.E. (1996) `Localization of the
photon on phase space', \emph{Journal of Mathematical Physics }\textbf{37},%
\textbf{\ }5958-5986.

\noindent Bub, J. (1997) \emph{Interpreting the Quantum World }(Cambridge:
Cambridge University Press).

\noindent Bub, J., and Clifton, R. (1996) `A Uniqueness Theorem for ``No
Collapse'' Interpretations of Quantum Mechanics', \emph{Studies in History
and Philosophy of Modern Physics }\textbf{27B}, 181-219.

\noindent Busch, P. (1985) `Indeterminacy Relations and Simultaneous
Measurements in Quantum Theory', \emph{International Journal of Theoretical
Physics }\textbf{24}, 63-92.

\noindent Busch, P. (1997) `Is the quantum state (an) observable?', in \emph{%
Potentiality, Entanglement and Passion-at-a-Distance}, Boston Studies in the
Philosophy of Science, Vol 194 (Boston: Kluwer).

\noindent Busch, P. (1998) `Can `unsharp objectification' solve the quantum
measurement problem?', \emph{International Journal of Theoretical Physics }%
\textbf{37}, 241-247.

\noindent Busch, P. (1999) `Unsharp localization and causality in
relativistic quantum theory', \emph{Journal of Physics A} \textbf{32},
6535-6546.

\noindent Busch, P., Grabowski, M. and Lahti, P. (1995) \emph{Operational
Quantum Physics }(Berlin: Springer), 2nd printing 1997.

\noindent Busch, P., and Lahti, P. (1996) `The Standard Model of Quantum
Measurement Theory: History and Applications', \emph{Foundations of Physics }%
\textbf{26}, 875-983.

\noindent Busch, P., Lahti, P., and Mittelstaedt, P. (1991) \emph{The
Quantum Theory of Measurement }(Berlin: Springer), 2nd revised edition 1996.

\noindent Busch, P., and Shimony, A. (1996) `Insolubility of the Quantum
Measurement Problem for Unsharp Observables', \emph{Studies in History and
Philosophy of Modern Physics }\textbf{27B}, 397-404.

\noindent Butterfield, J., and Fleming, G.N. (1999) `Strange Positions', in 
\emph{From Physics to Philosophy}, eds. J. Butterfield and C. Pagonis
(Cambridge: Cambridge University Press).

\noindent Cassinelli, G., De Vito, E., and Levrero, A. (1997) `On the
decompositions of a quantum state', \emph{Journal of Mathematical Analysis
and Applications}, \textbf{210}, 472-483.

\noindent Cologne (1998) \emph{Superluminal(?) Velocities: Tunneling time,
barrier penetration, non-trivial vacua, philosophy of physics}, Proceedings,
Cologne, June 1998, \emph{Annalen der Physik }\textbf{7}, 585-788.

\noindent Del Seta, M. (1998) \emph{Quantum Measurement as Theory: Its
Structure and Problems}, PhD Thesis, London School of Economics and
Political Science.

\noindent De Muynck, W.M., and Martens, H. (1990) `Neutron Interferometry
and the Joint Measurement of Incompatible Observables', \emph{Physical
Review A }\textbf{42}, 5079-5085.

\noindent De Muynck, W.M., Stoffels, W.W., and Martens, H. (1991) `Joint
Measurement of Interference and Path Observables in Optics and Neutron
Interferometry', \emph{Physica} \textbf{B 175}, 127-132.

\noindent Donald, M.J. (1999) `Book Review: Decoherence and the Appearance
of a Classical World in Quantum Theory', \emph{Studies in History and
Philosophy of Modern Physics }\textbf{30B}, 437-441.

\noindent D\H{u}rr, W., Nonn, T., and Rempe, G. (1998) `Origin of
quantum-mechanical complementarity probed by a `which-way' experiment in an
atom interferometer', \emph{Nature} \textbf{395 (3)}, 33-37.

\noindent Falkenburg, B. (1996) `The Analysis of Particle Tracks: A Case for
Trust in the Unity of Physics', \emph{Studies in History and Philosophy of
Modern Physics }\textbf{27B}, 337-371.

\noindent Fleming, G.N. (2000) `Essay Review: Operational Quantum Physics', 
\emph{Studies in History and Philosophy of Modern Physics }\textbf{31B},
117-125.

\noindent Giulini, D., Joos, E., Kiefer, C., Kupsch, J., Stamatescu, I.-O.,
and Zeh, H.D. (1996) \emph{Decoherence and the Appearance of a Classical
World in Quantum Theory }(Berlin: Springer).

\noindent Hadjisavvas, N. (1981) `Properties of mixtures of non-orthogonal
states' \emph{Letters on Mathematical Physics }\textbf{5}, 327-332.

\noindent Holland, P.R. (1993) \emph{The Quantum Theory of Motion -- An
Account of the de Broglie-Bohm Causal Interpretation of Quantum Mechanics }%
(Cambridge: Cambridge University Press).

\noindent Holland, S.S. (1995) `Orthomodularity in infinite dimensions; a
theorem of M. Sol\`{e}r', \emph{Bulletin of the American Mathematical Society%
} \textbf{32}, 205-234.

\noindent Hughston, L.P., Josza, R., and Wootters, W.K. (1993) `A complete
classification of quantum ensembles having a given density matrix', \emph{%
Physics Letters A }\textbf{183}, 14-18.

\noindent Lahti, P. (1980) `Uncertainty and complementarity in axiomatic
quantum mechanics', \emph{International Journal of Theoretical Physics }%
\textbf{19}, 789-842.

\noindent Lahti, P. (1983) `Hilbertian quantum theory as the theory of
complementarity', \emph{International Journal of Theoretical Physics }%
\textbf{22}, 911-929.

\noindent Lahti, P. (1987) `Complementarity and uncertainty: some
measurement-theoretical and information-theoretical aspects', in\emph{\
Symposium on the Foundations of Modern Physics 1997}, eds. P. Lahti and P.
Mittelstaedt, pp. 181-208 (Singapore: World Scientific).

\noindent Landsman, N.P. (1995) `Observation and Superselection in Quantum
Mechanics', \emph{Studies in History and Philosophy of Modern Physics }%
\textbf{26B}, 45-73.

\noindent Landsman, N.P. (1998) \emph{Mathematical topics between classical
and quantum mechanics} (New York: Springer).

\noindent Landsman, N.P. (1999) `Essay Review: Quantum Mechanics on Phase
Space', \emph{Studies in History and Philosophy of Modern Physics }\textbf{%
30B}, 282-305.

\noindent Lanz, L.(1994) `Quantum Physics and Objective Description', \emph{%
International Journal of Theoretical Physics }\textbf{33}, 19-29.

\noindent Lanz, L., and Melsheimer, O. (1993) `Towards an Objective Quantum
Physics', \emph{Nuovo Cimento B }\textbf{108}, 511-539.

\noindent L\'{e}vy-Leblond, J.-M. (1974) `The Pedagogical Role and
Epistemological Significance of Group Theory in Quantum Mechanics', \emph{%
Rivista del Nuovo Cimento} \textbf{4}, 99-143.

\noindent L\'{e}vy-Leblond, J.-M. (1981) `Classical apples and quantum
potatoes', \emph{European Journal of Physics }\textbf{2}, 44-47.

\noindent L\'{e}vy-Leblond, J.-M., and Balibar, F. (1990) \emph{Quantics --
Rudiments of Quantum Physics} (Amsterdam: North-Holland).

\noindent Ludwig, G. (1985) \emph{An Axiomatic Basis for Quantum Mechanics,
Vol 1: Derivation of Hilbert Space Structure }(Berlin: Springer).

\noindent Ludwig, G. (1987) \emph{An Axiomatic Basis for Quantum Mechanics,
Vol 2: Quantum Mechanics and Macrosystems }(Berlin: Springer).

\noindent Martens, H., and de Muynck, W.M. (1990a) `Nonideal Quantum
Measurements', \emph{Foundations of Physics}\textbf{\ 20}, 255-281.

\noindent Martens, H., and de Muynck, W.M. (1990b) `The Inaccuracy
Principle', \emph{Foundations of Physics}\textbf{\ 20}, 357-380.

\noindent Mittelstaedt, P. (1978) \emph{Quantum Logic }(Dordrecht: Reidel).

\noindent Mittelstaedt, P. (1986) \emph{Sprache und Realit\"{a}t in der
modernen Physik }(Mannheim: Bibliographisches Institut).

\noindent Mittelstaedt, P. (1995) `Constitution of Objects in Classical
Mechanics and in Quantum Mechanics', \emph{International Journal of
Theoretical Physics }\textbf{34}, 1615-1626.

\noindent Muga J.G., and Leavens, C.R. (2000) `Arrival Time in Quantum
Mechanics', \emph{Physics Reports}, to appear.

\noindent Ozawa, M. (1997) `Phase Operator problem and macroscopic extension
of quantum mechanics', \emph{Annals of Physics }\textbf{257}, 65-83.

\noindent Peres, A. (1993) \emph{Quantum Theory: Concepts and Methods }%
(Dordrecht: Kluwer).

\noindent Schr\"{o}dinger, E. (1936) `Probability relations between
separated systems', \emph{Proceedings of the Cambridge Philosphical Society }%
\textbf{32}, 446-452.

\noindent Schroeck, F.E. (1996) \emph{Quantum Mechanics on Phase Space }%
(Dordrecht: Kluwer).

\noindent Srinivas, M.D., and Vijayalakshmi, R.\ (1981) `The Time of
Occurrence in Quantum Mechanics', \emph{Pramana} \textbf{16}, 173-199.

\noindent Srinivas, M.D. (1996) `Quantum theory of continuous measurements
and its applications in quantum optics', \emph{Pramana} \textbf{47}, 1-23.

\noindent Stachow, E.W. (1980) `Logical Foundations of Quantum Mechanics', 
\emph{International Journal of Theoretical Physics }\textbf{19}, 251-304.

\noindent Stachow, E.W. (1985) `Structures of Quantum Language for
Individual Systems', in \emph{Recent Developments in Quantum Logic}, eds. P.
Mittelstaedt and E.W. Stachow (Mannheim: Bibliographisches Institut).

\noindent Strohmeyer, I. (1987) `Tragweite und Grenzen der
Transzendentalphilosophie zur Grundlegung der Quantenphysik', \emph{%
Zeitschrift f\H{u}r allgemeine Wissenschaftstheorie }\textbf{18/1-2}.

\noindent Strohmeyer, I. (1995) \emph{Quantentheorie und
Transzendentalphilosophie }(Heidelberg: Spektrum Akademischer Verlag).

\noindent Stulpe, W. (1997) \emph{Classical Representations of Quantum
Mechanics} (Berlin: Wissenschaft und Technik).

\noindent Uffink, J.B.M., and Hilgevoord, J. (1985) `Uncertainty Principle
and Uncertainty Relations' \emph{Foundations of Physics}\textbf{\ 15},
925-944.

\noindent Werner, R.F., and Wolff, M.P.H. (1995) `Classical Mechanics as
Quantum Mechanics with Infinitesimal $h$', \emph{Physics Letters A }\textbf{%
202}, 155-159.

\noindent Zapp, H. Chr. (1978) \emph{Die Spinbatterie -- ein Beitrag zur
quantenmechanischen Theorie des Messprozesses, }Diplomarbeit, Institute for
Theoretical Physics, University of Kiel, 1978.

\noindent Zbinden, H., Brendel, J., Gisin, N., and Tittel, W. (2000)
`Experimental test of non-local quantum correlations in relativistic
configurations', arXiv:quant-ph/0007009.

\end{document}